\documentclass[letterpaper, 10 pt, conference]{ieeeconf}

\IEEEoverridecommandlockouts                              

\usepackage{amsmath} 
\usepackage{amssymb}
\usepackage{hyperref}
\usepackage{mathtools}
\usepackage{optidef}
\usepackage{physics}
\usepackage{algorithm}
\usepackage[noend]{algpseudocode}
\usepackage{subcaption}
\usepackage{booktabs}
\usepackage{siunitx}

\let\oldproof\proof
\let\endoldproof\endproof
\let\proof\relax
\let\endproof\relax
\usepackage{amsthm}
\let\proof\oldproof
\let\endproof\endoldproof

\theoremstyle{definition}
\newtheorem{assumption}{Assumption}
\newtheorem{definition}{Definition}
\newtheorem{remark}{Remark}

\newlength{\Awidth}
\newlength{\Buwidth}
\newlength{\Bwwidth}
\newlength{\Cywidth}
\newlength{\Czwidth}
\newlength{\Dyuwidth}
\newlength{\Dywwidth}
\newlength{\Dzuwidth}
\newlength{\TxBwidth}
\newlength{\Dwidth}
\usepackage{fancyhdr}

\begin{document}

\title{\LARGE \bf
Inference and Learning of Nonlinear LFR State-Space Models*
}

\author{Merijn Floren$^{1,2}$, Jean-Philippe No{\"e}l$^{1}$, and Jan Swevers$^{1,2}$
\thanks{This work is supported by Flanders Make's IRVA project, ASSISStaNT. $^{1}$Department of Mechanical Engineering, KU Leuven, Belgium. $^{2}$Flanders Make@KU Leuven, Belgium. Corresponding author email: \texttt{merijn.floren@kuleuven.be}.}
}

\maketitle
\thispagestyle{empty}

\begin{abstract}
Estimating the parameters of nonlinear block-oriented state-space models from input-output data typically involves solving a highly non-convex optimization problem, which is prone to poor local minima and slow convergence. This paper presents a computationally efficient initialization method for nonlinear linear fractional representation (NL-LFR) models using periodic data.
By first inferring the latent signals and subsequently estimating the model parameters, the approach generates initial estimates for use in a later nonlinear optimization step. The proposed method shows robustness against poor local minima, and achieves a twofold error reduction compared to the state-of-the-art on a challenging benchmark dataset.
\end{abstract}


\section{Introduction}
The identification of nonlinear dynamical systems from input-output data is challenging primarily for two reasons: 
first, a suitable model structure must be selected to accurately represent the system's behavior; second, its parameters must be estimated, which is typically achieved through nonlinear optimization. Since the choice of model structure directly influences the parameter estimation process, it is essential to find the right balance between model expressiveness and the tractability of the resulting optimization problem.

\begin{figure}[t!]
    \centering
    \begin{minipage}{.485\textwidth}
        \centering
        \includegraphics[scale=1]{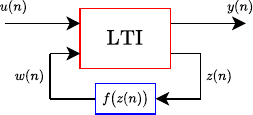}
        \subcaption{}
        \label{fig:LTIversion}
    \end{minipage}\\[1em]  
    \begin{minipage}{.485\textwidth}
        \centering
        \includegraphics[scale=1]{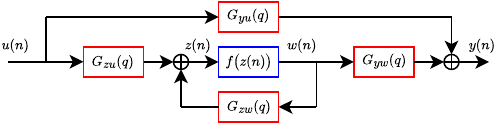}
        \subcaption{}
        \label{fig:TFversion}
    \end{minipage}
    \caption{The NL-LFR model as (a) a multi-input, multi-output LTI block with nonlinear feedback, and (b) an equivalent interaction between LTI blocks and a static nonlinearity.}
    \label{fig:model_structure}\vspace*{-16pt}
\end{figure}

State-space models are particularly appealing for modeling nonlinear systems, as they provide an efficient framework while also serving as the foundation for many modern model-based control techniques. A well-known example is the polynomial nonlinear state-space (PNLSS) model \cite{paduart2010identification}, which separates the nonlinear dynamics into a linear time-invariant (LTI) component and a static nonlinearity. Additional structure can be incorporated by recognizing that nonlinearities typically appear locally, motivating the use of block-oriented models \cite{giri2010block, schoukens2017identification}. Block-oriented models describe the interaction between LTI blocks and static nonlinearities, making them more interpretable while requiring fewer parameters than the more general PNLSS model. 
The focus of this work is on the nonlinear linear fractional representation (NL-LFR) model, which features a nonlinear feedback loop, essential for capturing the amplitude-dependent resonant behavior present in many mechanical systems. Any block-oriented model with a single static nonlinearity, e.g., the Hammerstein, Wiener, and Wiener-Hammerstein models, can be expressed in NL-LFR form, making it the most general model of this type. A visual representation of the model structure is shown in Fig. \ref{fig:model_structure}.

This paper focuses on NL-LFR \textit{simulation} models, which, unlike finite-horizon prediction models, present greater parameter estimation challenges as the relationship between the model parameters and the simulated output becomes more nonlinear with increasing data length. Consequently, the optimization problem is highly non-convex, making it susceptible to poor local minima and potentially slow convergence. 

One approach to reducing the complexity of the estimation problem is to initialize the NL-LFR model with the best linear approximation (BLA) \cite{pintelon2012system}, which can be parametrized non-iteratively using subspace methods \cite{mckelvey1996subspace}.
For weakly nonlinear systems, this significantly simplifies the nonlinear optimization problem, as only deviations from the BLA need to be learned.
Another approach to reducing the complexity of parameter estimation is to first infer the model's unmeasured signals  (\(w\) and \(z\) in Fig. \ref{fig:model_structure}), which allows for eliminating recursive dependencies and hence facilitates a more tractable optimization problem.
This strategy, known as inference and learning, is key to probabilistic methods such as the expectation-maximization algorithm \cite{turner2010state,schon2011system} and kernel-based methods leveraging canonical correlation analysis \cite{verdult2004least,van2009closed}. Indeed, inference and learning can be combined with initialization strategies, leading to both higher accuracy and reduced training times \cite{marconato2013improved,floren2024identification,floren2022nonlinear}.

Existing approaches for NL-LFR identification estimate the latent signals \(w\) and \(z\), but either rely on computationally expensive two-tone experiments \cite{vandersteen1999measurement}, or require BLA estimates at multiple operating conditions \cite{vanbeylen2013nonlinear}. Moreover, these methods are limited to cases where all five sub-blocks in Fig. \ref{fig:TFversion} are single-input single-output (SISO) systems, significantly restricting the modeling flexibility of the NL-LFR structure. Other works focus on improving initialization rather than inferring internal signals. In \cite{van2013identification}, a fully parametric NL-LFR initialization approach is proposed, though it remains limited to the SISO case. Meanwhile, \cite{schoukens2020initialization} imposes no restrictions on the dimensionality of the sub-blocks but relies solely on BLA-based initialization, which can still lead to slow convergence and suboptimal local minima. {Lastly, \cite{pavlov2013steady,shakib2022computationally,shakib2025numerical}, present a mixed-time-frequency identification approach to efficiently compute the steady-state response of NL-LFR models.}

This paper introduces a computationally efficient NL-LFR model identification method, combining BLA initialization with a novel time-frequency-domain inference and learning approach. The proposed method requires steady-state input-output data, and, as opposed to \cite{vandersteen1999measurement,vanbeylen2013nonlinear,van2013identification,pavlov2013steady,shakib2022computationally}, allows for arbitrarily-sized sub-blocks. Compared to \cite{van2013identification, schoukens2020initialization,pavlov2013steady,shakib2022computationally,shakib2025numerical}, {the latent signals are explicitly estimated to accelerate the learning of the nonlinearity. }
The remainder of this paper is structured as follows. Section \ref{sec:problem_statement} establishes the problem context, while Section \ref{sec:method} presents the proposed identification approach. Section \ref{sec:results} demonstrates the method's effectiveness on a challenging benchmark dataset, and Section \ref{sec:conclusions} concludes the paper.

\textit{Notation:} 
For a complex-valued vector \( x \in \mathbb{C}^n \) and a Hermitian, positive definite matrix \( Q \in \mathbb{C}^{n \times n} \), the squared weighted 2-norm of \( x \) is given by \(\|x\|_Q^2 = x^\mathrm{H} Q x\), where \( x^\mathrm{H} \) denotes the conjugate transpose of \( x \) (for a real-valued vector \( x \in \mathbb{R}^n \), the transpose is denoted by \( x^\top \)). The \( n \times n \) identity matrix is denoted by \( I_n \).
The discrete Fourier transform (DFT) of the time-domain samples \( \{x(n)\}_{n=0}^{N-1} \) at frequency index \( k \) is expressed as \( X(k) = \mathcal{F}_d\{x(n)\} \), while the inverse DFT (IDFT) at time instance \( n \), for the frequency-domain samples \( \{X(k)\}_{k=-N/2+1}^{N/2} \), is \( x(n) = \mathcal{F}_d^{-1}\{X(k)\} \).

\section{Problem statement}\label{sec:problem_statement}  
We consider the periodic in, same period out (PISPO) class, which contains systems where a periodic input produces a steady-state output with the same period.
\begin{definition}[Model representation]
The NL-LFR structure is parametrized as a discrete-time linear state-space model interconnected with a polynomial feedback nonlinearity:
\begin{subequations}\vspace{-4pt}
    \begin{align}
        x(n+1) &={A} x(n)+{B}_u u(n)+ B_w w(n),\label{eq:x_n}\\
        y_0(n) &= {C_y} x(n) + D_{yu} u(n) + D_{yw} w(n),\label{eq:y_n}\\
        z(n) &= C_z x(n) + D_{zu} u(n),\label{eq:z_n}\\
        w(n) &= \beta^\top \phi\big({z}(n)\big),\label{eq:w_n}
    \end{align}\label{eq:ss_model}
\end{subequations}
where \(x(n) \in \mathbb{R}^{n_x}\) represents the latent state at time index \( n \in \mathbb{Z} \), and \( u(n) \in \mathbb{R}^{n_u} \) and \( y_0(n) \in \mathbb{R}^{n_y} \) are the observable system inputs and outputs, respectively, while \( z(n) \in \mathbb{R}^{n_z} \) and \( w(n) \in \mathbb{R}^{n_w} \)  are the respective unobservable inputs and outputs of the static nonlinearity.
The matrices \( A \in \mathbb{R}^{n_x \times n_x} \), \( B_u \in \mathbb{R}^{n_x \times n_u} \), \( B_w \in \mathbb{R}^{n_x \times n_w} \), \( C_y \in \mathbb{R}^{n_y \times n_x} \), \( C_z \in \mathbb{R}^{n_z \times n_x} \), \( D_{yu} \in \mathbb{R}^{n_y \times n_u} \), \( D_{yw} \in \mathbb{R}^{n_y \times n_w} \), and \( D_{zu} \in \mathbb{R}^{n_z \times n_u} \) constitute the to-be-estimated LTI parameters. The static nonlinearity is modeled through \( \phi: \mathbb{R}^{n_z} \to \mathbb{R}^{n_\phi} \) and parameters \( \beta \in \mathbb{R}^{n_\phi \times n_w} \). The function \(\phi\) first applies a \(\tanh\) operation element-wise to its input \(z\), after which it maps the result to a feature vector of monomials, which may include cross-terms. \vspace{-4pt}
\end{definition}

\begin{remark}
    A direct feedthrough from \(w(n)\) to \(z(n)\) is not modeled, as this would result in an algebraic loop.\label{rem:direct_feedthrough}\vspace{-4pt}
\end{remark}
\begin{remark}\label{rem:tanh}
    {The \(\tanh\) operation acts as a smooth and differentiable saturation function that keeps the output elements of \(\phi\) bounded. In principle, any similarly well-behaved saturation function could be used. The rationale for introducing a saturation function is discussed in Section~\ref{sec:inference_learning}.}\vspace{-4pt}
\end{remark}
\begin{remark}
    {A polynomial basis is chosen due to its universal approximation capability for static nonlinear functions. Yet, any basis that is linear in the parameters could be used.instead. The polynomial degree and the inclusion of cross-terms are typically determined via cross-validation.}\vspace{-4pt}
\end{remark}

\begin{assumption}[Noise framework]
    The output measurements are corrupted by zero-mean, possibly colored, stationary noise \( v(n)\) with finite variance, i.e., 
    \(
        y(n) \coloneqq y_0(n) + v(n).
    \)
    The input \(u(n)\) is assumed to be noiseless.\vspace{-4pt}
\end{assumption}

Although the model dynamics (\ref{eq:ss_model}) are defined in the time domain, switching to the frequency domain enables a more efficient and parallelizable computation of the model response. 
This approach naturally aligns more with the block-oriented structure shown in Fig.~\ref{fig:TFversion}, composed of the LTI blocks:\vspace{-10pt}
\begin{equation}
    G_{ij}(\zeta) = C_i (\zeta I_{n_x} - A)^{-1} B_j + D_{ij},\label{eq:G_ij}
\end{equation}
for \(i=\{y,z\}\), \(j=\{u,w\}\), and where \( \zeta \) denotes the z-transform variable (\(q\) in Fig. \ref{fig:TFversion} represents the forward shift operator). Note that \(D_{zw}=0\), as per Remark \ref{rem:direct_feedthrough}.
    
The considered frequency-domain strategy requires periodic excitation signals, as formalized next.\vspace{-4pt}
\begin{definition}[Random-phase multisine]\label{def:multisine}
    The input excitations are described by a random-phase multisine signal \cite{pintelon2012system} with \(N\) samples per period, where \( N \) is even:\vspace{-4pt}
    \begin{equation}
        {u}(n) \coloneqq \frac{2}{\sqrt{N}} \sum_{k=1}^{N / 2 - 1} U_{k} \cos{(2\pi k f_0 n + \varphi_k)},
        \label{eq:multisine}\vspace{-4pt}
    \end{equation}
    where \( U_k \) denotes the excitation amplitude at frequency index \( k \), and \( f_0 = f_s/N \) is the frequency resolution, with \( f_s \) being the sampling frequency. The random phases $\varphi_k$ are independent and uniformly distributed on the interval \([0,2 \pi)\).\vspace{-4pt}
\end{definition}

{For parameter estimation, we consider \(P\) periods of \(R\) independent realizations of (\ref{eq:multisine}), resulting in dataset
\begin{equation}
    \mathcal{D} = \big\{\big(u^{\left[r\right]}(n), y^{\left[r\right]}(n)\big)\big\}_{n=0,r=0}^{N-1,R-1},
\end{equation}
where each signal represents the sample mean over the periods, i.e., \(u^{\left[r\right]}(n)=\frac{1}{P}\sum_{p=0}^{P-1}u^{\left[r,p\right]}(n)\) and \(y^{\left[r\right]}(n)=\frac{1}{P}\sum_{p=0}^{P-1}y^{\left[r,p\right]}(n)\). 
This averaging step reduces the impact of noise, as the variance of the sample means decreases inversely proportional with the number of periods \cite{pintelon2012system}.
We proceed by listing some additional conditions on \(\mathcal{D}\) that guarantee the validity or improve the quality of the model.\vspace{-4pt}
\begin{assumption}[Number of realizations]
    The number of independent realizations in \(\mathcal{D}\) must be at least as large as the number of inputs, i.e., \( R \geq n_u \), which ensures that the nonparametric BLA estimate is well-defined \cite{pintelon2012system}. Increasing the number of realizations enhances the richness of \(\mathcal{D}\), at the cost of greater computational complexity.\vspace{-4pt}
\end{assumption}}
\begin{assumption}[Steady state]\label{ass:steady_state}
    The dataset \(\mathcal{D}\) contains only steady-state signals. This is achieved by measuring multiple periods and discarding the ones that contain transients.\vspace{-4pt}
\end{assumption}
\begin{remark}
    {Definition~\ref{def:multisine} and Assumption~\ref{ass:steady_state} are essential to avoid spectral leakage, which otherwise introduces spurious frequency components that distort the parameter estimation process. If one wishes to use arbitrary, non-periodic, excitation data, it is important to reduce transient effects as much as possible by measuring over longer periods (i.e., increasing \(N\)) before taking the DFT \cite{pintelon2012system}. Addressing leakage explicitly, e.g., via the local polynomial method \cite{schoukens2009nonparametric}, remains a topic for future work.}\vspace{-4pt}
\end{remark}
\begin{assumption}[Standardization]\label{ass:standardization}
    For improved numerical stability and algorithm performance, the data in \(\mathcal{D}\) have been preprocessed to have zero mean and unit variance.\vspace{-4pt}
\end{assumption}

To estimate the model parameters \(\theta\), we minimize\vspace{-4pt}
\begin{equation}
    V(\theta) = \frac{1}{RN}\sum_{r=0}^{R-1} \sum_{k=0}^{N / 2} \big\|Y^{\left[r\right]}(k) - \hat{Y}^{\left[r\right]}(k,\theta)\big\|_{\Lambda(k)}^2, \label{eq:loss_function}\vspace{-4pt}
\end{equation}
where \( Y^{\left[r\right]}(k) =\mathcal{F}_d\{ y^{\left[r\right]}(n)\}\), and \( \hat{Y}^{\left[r\right]}(k,\theta) \) is the simulated output spectrum.
{The diagonal weighting matrix \(\Lambda(k)\) contains the inverses of the sample noise variances \cite{pintelon2012system}:\vspace{-4pt}
\begin{equation}
\sigma^2_y(k)=\frac{1}{R(P-1)}\sum_{r=0}^{R-             1}\sum_{p=0}^{P-1}|Y^{\left[r\right]}(k) - {Y}^{\left[r,p\right]}(k)|^2,
\label{eq:noise_variance}\vspace{-4pt}
\end{equation} 
where \( Y^{\left[r,p\right]}(k) =\mathcal{F}_d\{ y^{\left[r,p\right]}(n)\}\). In case \(P=1\), no noise estimate is available, so we set \(\Lambda(k) = I_{n_y}\).}
This work computes \(\hat{Y}^{\left[r\right]}(k,\theta)\) in (\ref{eq:loss_function}) in two ways: first via an inference and learning approach, and then using a more conventional forward simulation method. In both cases, (\ref{eq:loss_function}) is minimized using the Levenberg-Marquardt (LM) algorithm \cite{levenberg1944method}.

\section{Method}\label{sec:method}
We estimate the model parameters in three sequential steps. First, in Section \ref{sec:BLA}, \( \theta_{uy} =\{(A, B_u, C_y, D_{yu})\}\) is parametrized using the BLA. Second, in Section \ref{sec:inference_learning}, \(\theta_{wz}=\{(B_w, C_z, D_{yw}, D_{zu})\}\) and \( \beta \) are parametrized through a hybrid-domain inference and learning approach. Third, in Section \ref{sec:full_optimization}, all parameters \(\theta=\{\theta_{uy},\theta_{wz},\beta\}\) are jointly optimized in simulation mode.
\subsection{Best linear approximation}\label{sec:BLA}\vspace{-2pt}
Following \cite{pintelon2012system}, and given dataset \(\mathcal{D}\), the BLA of a nonlinear system within the PISPO class is defined as:
\begin{equation}
    \hat{G}(q) = \arg \min_{G} \mathbb{E}\left[ \|y(n) - G(q)u(n)\|^2 \right],\vspace{-4pt}
\end{equation}
where \(\mathbb{E}\left[\cdot\right]\) denotes the expected value operator.
In this work, we estimate the BLA in the frequency domain using the DFT of the input-output data. The procedure consists of three steps. First, a nonparametric estimate of \(\hat{G}\) is obtained (for details, see \cite{pintelon2012system}). Second, \(\hat{G}\) is parametrized, non-iteratively, in state-space form using the frequency-domain subspace method \cite{mckelvey1996subspace}. Third, the obtained parameters are refined through iterative LM optimization steps. The resulting matrices \(\hat{A}\), \(\hat{B}\), \(\hat{C}\), and \(\hat{D}\) are used to initialize:\vspace{-4pt}
\begin{equation}
    {\theta_{uy}}  =\{(T_x^{-1}\hat{A} T_x,\,\,
    T_x^{-1}\hat{B},\,\,
    \hat{C} T_x,\,\,
    \hat{D})\},
    \label{eq:theta_uy}\vspace{-4pt}
\end{equation}
where \(T_x\) is a diagonal matrix containing the standard deviations of the BLA state simulations. This similarity transform scales the linear state trajectories to have unit variance, which is beneficial for the subsequent optimization steps.

\subsection{Full nonlinear initialization}\label{sec:inference_learning}\vspace{-4pt}
The next step is to initialize all NL-LFR model parameters, which is achieved by minimizing (\ref{eq:loss_function}) with \(\theta_{wz}\) as decision variables and  \(\theta_{uy}\) as static ones. The optimization scheme follows an inference and learning approach, where \textit{inference} estimates the latent signals based on the observed data in \(\mathcal{D}\) and the LTI parameters, and \textit{learning} uses the inferred signals to efficiently estimate the coefficients \(\beta\).

Before optimization, \(\theta_{wz}\) is initialized using matrices \(B_w^*\), \(C_z^*\), \(D_{yw}^*\), and \(D_{zu}^*\), of which the elements are randomly sampled from a normal distribution with zero mean and unit variance. Specifically:\vspace{-4pt}
\begin{equation}
        {\theta}_{wz} = \{(B_w^*,\,\,
        T_z^{-1} C_z^*,\,\,
        {D}_{yw}^*,\,\,
        T_z^{-1} D_{zu}^* )\}.\label{eq:theta_init}\vspace{-4pt}
\end{equation}
For \(C_z\) and \(D_{zu}\) a scaling with diagonal matrix \(T_z\) is applied, of which the diagonal elements are defined as \((z_{\max}^* - z_{\min}^*)/2\). Here, \(z_{\min}^*,z_{\max}^* \in \mathbb{R}^{n_z}\) are the minimum and maximum values of the samples of \(z^*\), respectively, where \(z^*\) is obtained by simulating the BLA together with the initial values of \(C_z^*\) and \(D_{zu}^*\). This scaling approach ensures that the initial linear estimates of \(z\) are within the interval \(z\in[-1,1]\), thus providing a suitable input to the \(\tanh\) saturation function.
In the following, \(B_w^*\), \(C_z^*\), \(D_{yw}^*\), and \(D_{zu}^*\) are iteratively optimized while the BLA parameters are fixed. 

\subsubsection{Inference of hidden dynamics} 
The objective of this step is to obtain estimates of the unmeasured system dynamics. To do so, we formulate a frequency-domain optimization problem that minimizes the residual \(\mathcal{E}^{\left[r\right]}(k)={Y}^{\left[r\right]}(k) - \hat{Y}^{\left[r\right]}(k)\), while regularizing the required nonparametric input \(W^{\left[r\right]}(k)\):\vspace{-6pt}
\begin{equation}
    \min_{{W}}\sum_{r=0}^{R-1} \sum_{k=0}^{N/2} \big( \frac{1}{2}\|\mathcal{E}^{\left[r\right]}(k)\|^2_{\Lambda(k)} + \frac{\lambda}{2}\|{W}^{\left[r\right]}(k)\|^2_{\Theta}\big),\label{eq:inference}\vspace{-4pt}
\end{equation}
subject to
    \(
        \hat{Y}^{\left[r\right]}(k) = G_{yu}(\zeta_k){U}^{\left[r\right]}(k) + G_{yw}(\zeta_k){W}^{\left[r\right]}(k)\),
where \(\zeta_k\) is the z-transform variable evaluated on the unit circle at DFT frequency \(k\).
The hyperparameter \(\lambda \in \mathbb{R}_{>0}\) controls solution variability by penalizing the influence of \({W}^{\left[r\right]}(k)\) on the state and output through
\begin{equation}
    \Theta = \begin{bmatrix}
        B_w \\
        D_{yw}
    \end{bmatrix}^\top \begin{bmatrix}
        B_w\\
        D_{yw}
    \end{bmatrix} + \frac{\epsilon}{\lambda}I_{n_w},
\end{equation}
where \(\epsilon \in \mathbb{R}_{>0}\) is a small value that guarantees strict positive definiteness of \(\Theta\).{
Note that for \(\lambda\), numerically reasonable values (e.g., 0.01, 0.1, 1) should work well since both the states and outputs have been normalized (see (\ref{eq:theta_uy}) and Assumption~\ref{ass:standardization}).}
The optimization problem (\ref{eq:inference}) is convex for every \(k\) and \(r\), with closed-form solution:\vspace{-4pt}
\begin{equation}
    {W}_*^{\left[r\right]}(k) = \Omega^{-1}\Psi \big({Y}^{\left[r\right]}(k) - G_{yu}(\zeta_k){U}^{\left[r\right]}(k)\big),\label{eq:W_star}
\end{equation}
where \(\Psi = G^\mathrm{H}_{yw}(\zeta_k)\Lambda(k)\), and \(\Omega=\Psi G_{yw}(\zeta_k) + \lambda \Theta\).
Next, the corresponding estimate of \({Z}^{\left[r\right]}(k)\) is computed as
\begin{equation}
    {Z}_*^{\left[r\right]}(k) = G_{zu}(\zeta_k){U}^{\left[r\right]}(k) + G_{zw}(\zeta_k){W}_*^{\left[r\right]}(k), \label{eq:Z_star}\vspace{-4pt}
\end{equation}
thereby concluding the inference step.
\subsubsection{Learning of model parameters}
The process of learning consists of three consecutive stages, outlined as follows.
\begin{figure}
    \centering
    \includegraphics[width=\columnwidth]{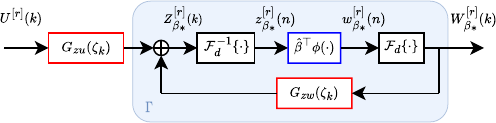}
    \caption{Schematic of a consistent feedback loop: fixed point \({Z}_{\beta_*}^{\left[r\right]}(k)\) is invariant to the feedback transformation \(\Gamma\).}
    \label{fig:feedback_consistency}\vspace*{-16pt}
\end{figure}

\textit{(\romannumeral1) Solving for  \(\beta\):} Using the obtained nonparametric spectral estimates \({W}_*^{\left[r\right]}(k)\) and \({Z}_*^{\left[r\right]}(k)\), the coefficients \(\beta\) can be learned by solving an estimation problem that is linear in the parameters. We perform this step in the time domain, since this significantly simplifies the construction of the nonlinear monomials. By defining \(\mathbf{w}_* \in \mathbb{R}^{(NR) \times n_w}\) and \(\mathbf{z}_* \in \mathbb{R}^{(NR) \times n_z}\) as the stacked IDFTs of \({W}_*^{\left[r\right]}(k)\) and \({Z}_*^{\left[r\right]}(k)\), respectively, we can express the nonlinear relationship in (\ref{eq:w_n})
in matrix-form as \(\mathbf{w}^*=\Phi({\mathbf{z}_*})\beta\), where \(\Phi: \mathbb{R}^{(NR)\times n_z} \to \mathbb{R}^{(NR) \times n_\phi}\) applies \(\phi\) row-wise to its input. 
The estimation of \(\beta\) then simplifies to:\vspace{-4pt}
\begin{equation}
    \hat{\beta} = \big(\Phi^\top({\mathbf{z}}_*)\Phi({\mathbf{z}}_*)\big)^{-1}\Phi^\top({\mathbf{z}}_*)\mathbf{w}_*,\label{eq:beta}\vspace{-4pt}
\end{equation}
which, together with \({\theta_{uy}}\) and \({\theta_{wz}}\), fully parametrizes the NL-LFR model.

\textit{(\romannumeral2) Promoting internal consistency:}
The estimation of \(\beta\) in (\ref{eq:beta}) is simple and efficient but implicitly assumes that the samples in \(\mathbf{z}_*\) and \(\mathbf{w}_*\) are independently and identically distributed.
This assumption breaks down during forward simulation of the dynamics, as the output \(w\) of the nonlinearity \(\phi\) at time \(n\) influences its own input \(z\) at time \(n+1\). Consequently, the residual error of the estimate \(\hat{\beta}\) no longer remains locally contained but instead accumulates over time, causing a mismatch between the estimation and simulation environments that may quickly lead to divergence \cite{venkatraman2015improving}.

The risk of divergence necessitates accounting for internal consistency within the optimization problem. Specifically, the parametric output spectrum \(\hat{Y}^{[r]}(k,\theta)\) in (\ref{eq:loss_function}) must be computed from a parameter set \(\theta\) that yields stable time-domain behavior. To formalize this requirement, we define the dynamics of a consistent feedback loop as:\vspace{-4pt}
\begin{equation}
    \mathbf{Z}_{\beta_*}^{\left[r\right]} = \Gamma\big(\mathbf{Z}_{\beta_*}^{\left[r\right]}\big),\label{eq:Z_implicit}\vspace{-4pt}
\end{equation}
where \(\Gamma: \mathbb{C}^{N \times n_z} \to \mathbb{C}^{N \times n_z}\) represents the implicit feedback operator that maps the matrix \(\mathbf{Z}_{\beta_*}^{\left[r\right]} \in \mathbb{C}^{N \times n_z}\) onto itself. In other words, for the internal dynamics to be consistent, \(\mathbf{Z}_{\beta_*}^{\left[r\right]}\) must be a fixed point of \(\Gamma\), as illustrated in Fig. \ref{fig:feedback_consistency}. However, the matrix \(\mathbf{Z}_{*}^{\left[r\right]}\in \mathbb{C}^{N \times n_z}\) (which contains \({Z}_*^{\left[r\right]}(k)\) for all \(k\)) in combination with the imperfect estimate \(\hat{\beta}\), generally fails to meet the fixed-point condition, and should hence not be used naively in the optimization problem. Still, \(\mathbf{Z}_{*}^{\left[r\right]}\) can be used to initialize an iterative fixed-point search, as formalized next.\vspace{-4pt}


\begin{algorithm}[b]
    \caption{Inference and learning of NL-LFR models}\label{alg:inference_learning}
    \begin{algorithmic}[1]
        \State \textbf{Input:} Dataset \(\mathcal{D}\), BLA parameters \(\theta_{uy}\), regularization parameter \(\lambda\), and number of fixed-point iterations \(\tau\)
        \State Initialize \(\theta_{wz}\) using random values \Comment{Eq. (\ref{eq:theta_init})}
        \While{not converged}
        \State \textbf{1) Inference of hidden dynamics}
        \State Compute \({W}_*^{\left[r\right]}(k),\:{Z}_*^{\left[r\right]}(k)\) \(\;\forall\, k, r
\) \Comment{Eqs. (\ref{eq:W_star})--(\ref{eq:Z_star})}
        \State \textbf{2) Learning of model parameters}
        \State (\romannumeral1) Solve a linear problem to obtain \(\hat{\beta}\) \Comment{Eq. (\ref{eq:beta})}
        \State (\romannumeral2) Perform \(\tau\) fixed-point iterations \Comment{Eq. (\ref{eq:fixed_point_iter})}
        \State (\romannumeral3) Update elements of \(\theta_{wz}\) \Comment{Eqs. (\ref{eq:Y_par}) and (\ref{eq:loss_function})}
        \EndWhile
        \State \textbf{Output:} Fully initialized NL-LFR model
    \end{algorithmic}
\end{algorithm}
\begin{definition}[Fixed-point iterations]\label{def:fixed_point}
    Given the nonparametric estimate \(\mathbf{Z}_{\beta_0}^{\left[r\right]} = \mathbf{Z}_{*}^{\left[r\right]}\), we perform \(\tau \in  \mathbb{Z}_{> 0}\) fixed-point iterations:\vspace{-4pt}
    \begin{equation}
        \big\{\mathbf{Z}_{\beta_{i+1}}^{\left[r\right]} = \Gamma\big(\mathbf{Z}_{\beta_{i}}^{\left[r\right]}\big)\big\}_{i=0}^{\tau-1},\label{eq:fixed_point_iter}\vspace{-4pt}
    \end{equation}
   {which are expected to converge to a consistent solution.}\vspace{-4pt}
\end{definition}
\begin{remark}
    It is generally not required for \(\mathbf{Z}_{\beta_{\tau}}^{\left[r\right]}\) to be an \textit{exact} fixed point of \(\Gamma\). Instead, it needs to be sufficiently close to ensure bounded and numerically stable time-domain NL-LFR simulations.{ Thanks to the careful nonlinear initialization of the fixed-point scheme, it is expected that a relatively low number of iterations (e.g., \(\tau=3\))  typically suffices to ensure stable time-domain behavior.}\vspace{-4pt}
\end{remark}

{The fixed-point iterations in Definition~\ref{def:fixed_point} closely resemble the mixed-time-frequency (MTF) approach proposed in \cite{pavlov2013steady,shakib2022computationally,shakib2025numerical}, which constrains the parameters \(\theta\) to be within a globally exponentially convergent set. By doing so, convergence to a unique fixed point is guaranteed, independent of the initial condition. 
However, the enforced conditions are only sufficient for convergence and thus reject many other valid solutions, leading to conservative estimates that significantly reduce model performance compared to the unconstrained case \cite{shakib2025numerical}.}

{To maximize model accuracy and leverage the efficient unconstrained LM optimization algorithm, we deliberately avoid enforcing such convergence conditions. This choice can potentially lead to (i) convergence to a stable but non-unique fixed point, or (ii) diverging fixed-point iterations.
An implication of the first scenario is that the inference and learning approach could compute a different steady-state response than the time-domain simulations, as they start from different initial conditions. Regarding the second scenario, we argue that stability of the iterations is promoted by the \(\tanh\) saturation function that provides a finite upper bound on the incremental sector condition of \(\phi\), which is one of the MTF convergence criteria. The effectiveness of the proposed fixed-point iterations in terms of time-domain stability and performance is experimentally validated in Section \ref{sec:results}.}

\textit{(\romannumeral3) Updating \(\theta_{wz}\):} 
Based on the final fixed-point iteration, we extract the samples \({Z}_{\beta_{\tau}}^{\left[r\right]}(k)\) from \(\mathbf{Z}_{\beta_{\tau}}^{\left[r\right]}\) to compute\vspace{-4pt}
\begin{equation}
    \hat{Y}^{\left[r\right]}(k,\theta) = G_{yu}(\zeta_k){U}^{\left[r\right]}(k) + G_{yw}(\zeta_k){W}_{\beta_{\tau}}^{\left[r\right]}(k),\label{eq:Y_par}\vspace{-2pt}
\end{equation}
where
\(
    {W}_{\beta_{\tau}}^{\left[r\right]}(k) = \mathcal{F}_d\big\{\hat{\beta}^\top \phi\big(\mathcal{F}_d^{-1}\{{Z}_{\beta_{\tau}}^{\left[r\right]}(k)\}\big)\big\}.
\)
The LM algorithm then iteratively adjust the decision variables \(\theta_{wz}\) based on the loss in (\ref{eq:loss_function}), continuing until either convergence is achieved or the maximum number of iterations is reached.
A complete overview of the proposed inference and learning approach is provided in Algorithm \ref{alg:inference_learning}.\vspace{-6pt}

\subsection{Full nonlinear optimization}\label{sec:full_optimization}\vspace{-4pt}
At this stage, all NL-LFR parameters have been initialized, but further optimization is needed to improve the simulation performance of the model.
Therefore, we optimize all parameters \(\theta\) using forward simulations driven by \(
u^{\left[r\right]}(n)\). We define \(\hat{Y}^{\left[r\right]}(k,\theta)\) as the DFT of the simulated time-domain output (ensuring it reached steady state) to minimize (\ref{eq:loss_function}). 
It may be expected that poor local minima are avoided since the previous steps should have already provided a good initialization.\vspace{-10pt}

\section{Results}\label{sec:results}\vspace{-4pt}
We evaluate the effectiveness of the proposed method on the parallel Wiener-Hammerstein benchmark system \cite{schoukens2015parametric}, where each of the two branches contains a diode-resistor static nonlinearity, sandwiched between two third-order linear filters.
The overall SISO system exhibits 12th-order dynamics (\(n_x=12\)) and can be reformulated as an NL-LFR structure with a multi-input multi-output (MIMO) nonlinearity (\(n_w=n_z=2\)). The nonlinearity is modeled using monomial basis functions up to degree 7, including cross-terms. The total number of model parameters is 293.

For estimation, we use \(P=2\) periods of \(R=5\) multisine realizations with \(N=16384\) samples at a root-mean-square (RMS) excitation amplitude of \SI{1}{\volt}.
All experiments are conducted on an NVIDIA GeForce RTX 2080 Ti GPU. Further implementation details are provided in this work's repository: \href{https://github.com/merijnfloren/sysid-nonlinear-lfr}{\textcolor{cyan}{https://github.com/merijnfloren/sysid-nonlinear-lfr}}.

\begin{figure}[t]
    \centering
    \includegraphics[width=.485\textwidth]{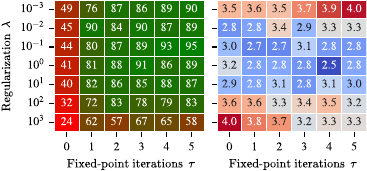}
    \caption{
        Impact of \(\tau\) and \(\lambda\) on the time-domain performance of the inference and learning method, based on 100 random initializations per grid point: (left) number of runs that led to improved time-domain performance; (right) corresponding simulation error (averaged, and expressed as a percentage).
    }
    \label{fig:hyperparameter}
\end{figure}

First, we assess the effect of the fixed-point iterations and the regularization in (\ref{eq:inference}) on the time-domain stability of the inference and learning method.
Specifically, we perform a grid search over \(\lambda \in \{10^{-3},10^{-2},\ldots,10^3\}\) and \(\tau \in \{0,1,\ldots,5\}\). For each combination we run 250 LM iterations of inference and learning, and we repeat each trial 100 times with different random initializations of \(\theta_{wz}\).
The results are shown in Fig. \ref{fig:hyperparameter}, where the left grid indicates the number of trials that resulted in improved (w.r.t. the BLA) time-domain simulations; the right grid presents the corresponding mean relative simulation errors as percentages (for reference, the BLA error is \SI{11.18}{\percent}). Fig. \ref{fig:hyperparameter} shows that the success rate generally increases with \(\tau\), which aligns with expectations, as higher \(\tau\) values enhance the model's internal consistency and hence reduce accumulating time-domain errors. {Note that unsuccessful trials either indicate instability from accumulating errors caused by insufficient fixed-point convergence, or reflect a mismatch with the inference and learning approach due to differing initial conditions.}
Moreover, the initialization method demonstrates robustness with respect to \(\lambda\): a broad range of values leads to improved simulations with low errors; the performance only notably deteriorates at the extremes, both in terms of successful trials and simulation accuracy. In summary, inference and learning achieves a significant improvement over the BLA, and is generally stable across hyperparameters.
\begin{figure}[t]
    \centering
    \includegraphics[width=.485\textwidth]{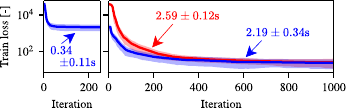}
    \caption{
        Training progress of 100 random initializations of \(\theta_{wz}\): (left) inference and learning method for \(\lambda = 1\) and \(\tau = 3\); (right) full nonlinear optimization. Blue: proposed method; red: baseline method. The embedded numbers indicate the computation time per iteration (mean \(\pm\) standard deviation).
    }
    \label{fig:progress}\vspace*{-22pt}
\end{figure}

Next, we examine the impact of the inference and learning method on the final optimization stage. We do so by comparing the full proposed method to a baseline approach where the final optimization is initialized merely with the BLA. For the proposed method, we first perform 250 LM iterations of inference and learning with \(\lambda=1\) and \(\tau=3\), followed by 1000 LM iterations of nonlinear optimization. For the baseline approach, we directly perform 1000 LM iterations of nonlinear optimization.
Again, each trial is repeated 100 times, with both methods starting from the same randomly initialized \(\theta_{wz}\) in every run. For the baseline method, the elements of \(\beta\) are initially set to zero, ensuring that the initial loss matches the BLA loss. Fig. \ref{fig:progress} shows the median of the respective training loss progressions for all 100 trials, with shaded regions indicating the median absolute deviation.
The left plot in Fig. \ref{fig:progress} shows that inference and learning quickly reduces the training loss with roughly a factor of 20, after which it plateaus, likely due to the optimizer's degrees of freedom being restricted to \(\theta_{wz}\) alone. From the right plot of Fig. \ref{fig:progress} it can be seen that the obtained initial estimates provide a head start for the final optimization. As a result, the proposed method converges faster and to a slightly lower minimum compared to the baseline method.
Notably, inference and learning produced 91 stable initializations, all of which converged to significantly lower local minima during final optimization. In contrast, only 85 trials of the baseline method showed significant improvement over the BLA, which suggests that starting further from a good local minimum made them more prone to getting stuck in poor local minima. Fig. \ref{fig:progress} also reports on the respective computational times per iteration, where, thanks to its parallelizable nature, the inference and learning iterations are over six times faster than those of the full nonlinear optimization.
\begin{figure}[t]
    \centering
    \includegraphics[width=.485\textwidth]{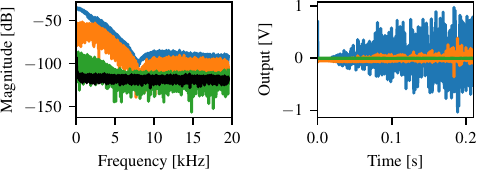}
      \caption{Test simulations: (left) multisine in frequency domain; (right) arrowhead in time domain. Blue: system output; orange: BLA error; green: NL-LFR error; {black: noise level}.}
    \label{fig:test}\vspace{-4pt}
\end{figure}
\begin{table}[t]
    \caption{Comparison of simulation RMS errors (in volts) on the multisine and arrowhead test datasets.}
    \label{tab:rmses}
    \centering
    {\fontsize{8.45}{8.45}\selectfont
    \begin{tabular}{l|cccc}
           & BLA & Proposed   & ParWH \cite{schoukens2015parametric}  & NL-LFR \cite{schoukens2020initialization}  \\
    \midrule
    Multisine  &  3.77e$^{-2}$  & 6.04e$^{-4}$  & 1.10e$^{-3}$    & 1.37e$^{-3}$    \\
    Arrowhead  & 4.54e$^{-2}$ &  6.12e$^{-4}$  & 2.66e$^{-3}$   & 1.42e$^{-3}$   \\
    \end{tabular}
    }\vspace*{-12pt}
\end{table}

Finally, from the trials in Fig. \ref{fig:progress}, we select the model with the lowest training loss and test it on two datasets: (i) another multisine realization with the same properties as the estimation data, and (ii) a Gaussian noise sequence with a linearly increasing amplitude over time (denoted as the arrowhead signal).
The simulation results are shown in Fig. \ref{fig:test}, where it is clear that the NL-LFR model significantly outperforms the BLA. {Moreover, the multisine plot shows that the NL-LFR error closely follows the noise floor \(\sigma_y(k)\), which was derived from the training data via (\ref{eq:noise_variance}).} The corresponding relative simulation errors are reduced from \SI{11.29}{\percent} to \SI{0.18}{\percent} for the multisine data, and from \SI{17.36}{\percent} to \SI{0.23}{\percent} for the arrowhead signal. This improvement is also evident from Table \ref{tab:rmses}, where the proposed method is compared to the BLA and two state-of-the-art approaches: the first method explicitly enforces the parallel Wiener-Hammerstein structure \cite{schoukens2015parametric}, while the second method initializes an NL-LFR model with the BLA and then models the residual \cite{schoukens2020initialization} (similar to the baseline method). The selected model of this work outperforms both methods on both test datasets, achieving approximately a twofold RMS error reduction, while requiring a similar number of parameters.\vspace{-8pt}

\section{Conclusions}\label{sec:conclusions}\vspace{-8pt}
This paper presented a novel method for identifying NL-LFR state-space models from input-output data. By combining BLA-based initialization with latent signal inference and nonlinear parameter learning, the proposed method reduces susceptibility to poor local minima and achieves the lowest error on a challenging benchmark dataset. Additionally, the method's inherent parallelism allows for quickly exploring multiple model architectures, making it a practical tool for nonlinear system identification.
{
Future work includes extending the algorithm to non-periodic data, and introducing a more flexible notion of stability, e.g., treating  global convergence conditions as soft constraints or formulating an alternative, less conservative set of convergence criteria.
}

\bibliographystyle{IEEEtran}
\bibliography{IEEEabrv, references}

\begin{thebibliography}{10}
\providecommand{\url}[1]{#1}
\csname url@rmstyle\endcsname
\providecommand{\newblock}{\relax}
\providecommand{\bibinfo}[2]{#2}
\providecommand\BIBentrySTDinterwordspacing{\spaceskip=0pt\relax}
\providecommand\BIBentryALTinterwordstretchfactor{4}
\providecommand\BIBentryALTinterwordspacing{\spaceskip=\fontdimen2\font plus
\BIBentryALTinterwordstretchfactor\fontdimen3\font minus
  \fontdimen4\font\relax}
\providecommand\BIBforeignlanguage[2]{{%
\expandafter\ifx\csname l@#1\endcsname\relax
\typeout{** WARNING: IEEEtran.bst: No hyphenation pattern has been}%
\typeout{** loaded for the language `#1'. Using the pattern for}%
\typeout{** the default language instead.}%
\else
\language=\csname l@#1\endcsname
\fi
#2}}

\bibitem{paduart2010identification}
J.~Paduart, L.~Lauwers, J.~Swevers, K.~Smolders, J.~Schoukens, and R.~Pintelon,
  ``Identification of nonlinear systems using polynomial nonlinear state space
  models,'' \emph{Automatica}, vol.~46, no.~4, 2010.

\bibitem{giri2010block}
F.~Giri and E.-W. Bai, \emph{Block-oriented nonlinear system
  identification}.\hskip 1em plus 0.5em minus 0.4em\relax Springer, 2010,
  vol.~1.

\bibitem{schoukens2017identification}
M.~Schoukens and K.~Tiels, ``Identification of block-oriented nonlinear systems
  starting from linear approximations: A survey,'' \emph{Automatica}, vol.~85,
  pp. 272--292, 2017.

\bibitem{pintelon2012system}
R.~Pintelon and J.~Schoukens, \emph{System identification: a frequency domain
  approach}.\hskip 1em plus 0.5em minus 0.4em\relax John Wiley \& Sons, 2012.

\bibitem{mckelvey1996subspace}
T.~McKelvey, H.~Ak{\c{c}}ay, and L.~Ljung, ``Subspace-based multivariable
  system identification from frequency response data,'' \emph{IEEE Transactions
  on Automatic Control}, vol.~41, no.~7, pp. 960--979, 1996.

\bibitem{turner2010state}
R.~Turner, M.~Deisenroth, and C.~Rasmussen, ``State-space inference and
  learning with {G}aussian processes,'' in \emph{Proceedings of the Thirteenth
  International Conference on Artificial Intelligence and Statistics}.\hskip
  1em plus 0.5em minus 0.4em\relax JMLR Workshop and Conference Proceedings,
  2010.

\bibitem{schon2011system}
T.~B. Sch{\"o}n, A.~Wills, and B.~Ninness, ``System identification of nonlinear
  state-space models,'' \emph{Automatica}, vol.~47, no.~1, 2011.

\bibitem{verdult2004least}
V.~Verdult, J.~A. Suykens, J.~Boets, I.~Goethals, and B.~De~Moor, ``Least
  squares support vector machines for kernel {C}{C}{A} in nonlinear state-space
  identification,'' in \emph{Proceedings of the 16th International Symposium on
  Mathematical Theory of Networks and Systems (MTNS2004), Leuven, Belgium},
  2004.

\bibitem{van2009closed}
J.-W. van Wingerden and M.~Verhaegen, ``Closed-loop subspace identification of
  {H}ammerstein-{W}iener models,'' in \emph{Proceedings of the 48h IEEE
  Conference on Decision and Control (CDC) held jointly with 2009 28th Chinese
  Control Conference}.\hskip 1em plus 0.5em minus 0.4em\relax IEEE, 2009, pp.
  3637--3642.

\bibitem{marconato2013improved}
A.~Marconato, J.~Sj{\"o}berg, J.~A. Suykens, and J.~Schoukens, ``Improved
  initialization for nonlinear state-space modeling,'' \emph{IEEE Transactions
  on Instrumentation and Measurement}, vol.~63, no.~4, 2013.

\bibitem{floren2024identification}
M.~Floren, S.~Mamedov, J.-P. No{\"e}l, and J.~Swevers, ``Identification of
  {D}eformable {L}inear {O}bject {D}ynamics from {I}nput-output {M}easurements
  in 3{D} {S}pace,'' \emph{IFAC-PapersOnLine}, vol.~58, no.~15, pp. 468--473,
  2024.

\bibitem{floren2022nonlinear}
M.~Floren and J.-P. No{\"e}l, ``Nonlinear restoring force modelling using
  {G}aussian processes and model predictive control,'' in \emph{Conference
  Proceedings of ISMA2022-USD2022}, 2022, pp. 2493--2498.

\bibitem{vandersteen1999measurement}
G.~Vandersteen and J.~Schoukens, ``Measurement and identification of nonlinear
  systems consisting of linear dynamic blocks and one static nonlinearity,''
  \emph{IEEE Transactions on Automatic Control}, vol.~44, no.~6, pp.
  1266--1271, 1999.

\bibitem{vanbeylen2013nonlinear}
L.~Vanbeylen, ``Nonlinear {L}{F}{R} block-oriented model: {P}otential benefits
  and improved, user-friendly identification method,'' \emph{IEEE Transactions
  on Instrumentation and Measurement}, vol.~62, no.~12, 2013.

\bibitem{van2013identification}
A.~Van~Mulders, J.~Schoukens, and L.~Vanbeylen, ``Identification of systems
  with localised nonlinearity: {F}rom state-space to block-structured models,''
  \emph{Automatica}, vol.~49, no.~5, pp. 1392--1396, 2013.

\bibitem{schoukens2020initialization}
M.~Schoukens and R.~T{\'o}th, ``On the initialization of nonlinear {L}{F}{R}
  model identification with the best linear approximation,''
  \emph{IFAC-PapersOnLine}, vol.~53, no.~2, pp. 310--315, 2020.

\bibitem{pavlov2013steady}
A.~Pavlov, B.~Hunnekens, N.~van~de Wouw, and H.~Nijmeijer, ``Steady-state
  performance optimization for nonlinear control systems of {L}ur’e type,''
  \emph{Automatica}, vol.~49, no.~7, pp. 2087--2097, 2013.

\bibitem{shakib2022computationally}
M.~F. Shakib, A.~Y. Pogromsky, A.~Pavlov, and N.~van~de Wouw, ``Computationally
  efficient identification of continuous-time {L}ur’e-type systems with
  stability guarantees,'' \emph{Automatica}, vol. 136, p. 110012, 2022.

\bibitem{shakib2025numerical}
M.~F. Shakib, N.~Vervaet, A.~S. Pogromsky, A.~Pavlov, and N.~van~de Wouw,
  ``Numerical tools for the efficient steady-state optimization of
  discrete-time {L}ur'e-type control systems,'' in \emph{13th IFAC Symposium on
  Nonlinear Control Systems}, 2025.

\bibitem{schoukens2009nonparametric}
J.~Schoukens, G.~Vandersteen, K.~Barb{\'e}, and R.~Pintelon, ``Nonparametric
  preprocessing in system identification: a powerful tool,'' in \emph{2009
  European Control Conference (ECC)}.\hskip 1em plus 0.5em minus 0.4em\relax
  IEEE, 2009, pp. 1--14.

\bibitem{levenberg1944method}
K.~Levenberg, ``A method for the solution of certain non-linear problems in
  least squares,'' \emph{Quarterly of applied mathematics}, vol.~2, no.~2, pp.
  164--168, 1944.

\bibitem{venkatraman2015improving}
A.~Venkatraman, M.~Hebert, and J.~Bagnell, ``Improving multi-step prediction of
  learned time series models,'' in \emph{Proceedings of the AAAI Conference on
  Artificial Intelligence}, vol.~29, no.~1, 2015.

\bibitem{schoukens2015parametric}
M.~Schoukens, A.~Marconato, R.~Pintelon, G.~Vandersteen, and Y.~Rolain,
  ``Parametric identification of parallel {W}iener--{H}ammerstein systems,''
  \emph{Automatica}, vol.~51, pp. 111--122, 2015.

\end{thebibliography}

\end{document}